\documentclass[12pt,british]{article}

\usepackage[utf8]{inputenc}
\usepackage{geometry}
\usepackage{float}
\usepackage{booktabs}
\usepackage{multirow}
\usepackage{amsmath}
\usepackage{amsthm}
\usepackage{amssymb}
\usepackage{graphicx}
\usepackage{caption}
\usepackage{subcaption}
\usepackage{amsthm}
\usepackage{braket}
\usepackage{multicol}
\usepackage{color}
\usepackage{textcase}
\usepackage{setspace}
\usepackage{fancyhdr}
\usepackage{enumerate}
\usepackage{url}
\usepackage{babel}
\usepackage{cite}

\numberwithin{equation}{section}
\numberwithin{figure}{section}
\numberwithin{table}{section}

\makeatletter

\newcommand{\op}{\mathcal{O}}

\newcommand{\e}[1]{e^{#1}}

\makeatother

\begin{document}
\title{Weak integrability breaking and level spacing distribution}
\author{D. Szász-Schagrin$^{1}$, B. Pozsgay$^{2}$ and G. Takács$^{1,3}$ \\
$^{1}${\small{}{}{}BME Momentum Statistical Field Theory Research
Group},\\
 {\small{}{}{}Department of Theoretical Physics,}\\
 {\small{}{}{}Budapest University of Technology and Economics}\\
 $^{2}${\small{}{}{}MTA-ELTE “Momentum” Integrable Quantum Dynamics
Research Group (ELKH),}\\
 {\small{}{}{}Department of Theoretical Physics, Eötvös Loránd University,
Budapest}\\
  $^{3}${\small{}{}{}MTA-BME Quantum Correlations Group (ELKH),}\\
 {\small{}{}{}Department of Theoretical Physics,}\\
 {\small{}{}{}Budapest University of Technology and Economics} }
\date{19th August 2023}

\maketitle

\begin{abstract}
Recently it was suggested that certain perturbations of integrable
spin chains lead to a weak breaking of integrability in the sense
that integrability is preserved at the first order in the coupling.
Here we examine this claim using level spacing distribution. We find
that the volume dependent crossover between integrable and chaotic level spacing statistics {which marks the onset of quantum chaotic behaviour, is markedly different for weak vs. strong breaking of integrability. In particular}, for the gapless case we find that the crossover coupling as a function of the volume $L$ scales with a $1/L^{2}$ law for weak breaking
as opposed to the $1/L^{3}$ law previously found for the strong case. 
\end{abstract}

\section{Introduction}

Mean values of generalised currents in one-dimensional integrable
models have attracted considerable interest lately, mainly due to
the recent theory of Generalised Hydrodynamics which describes non-equilibrium
dynamics at the Euler scale \cite{2016PhRvX...6d1065C,2016PhRvL.117t7201B}.
These currents express the continuity relation for conserved charges
responsible for integrability, which can be exploited for a hydrodynamic
description of the ballistic flow of the quasi-particles. This description
is useful only as far as the exact mean values of the currents in
local equilibrium conditions are known, and indeed in the thermodynamic
limit a simple exact formula was postulated in \cite{2016PhRvX...6d1065C,2016PhRvL.117t7201B},
which was proven for relativistic quantum field theories \cite{2016PhRvX...6d1065C,2019ScPP....6...23V},
and for the spin current of the XXZ spin chain in \cite{2019ScPP....6....5U},
and the classical Toda chain in \cite{2019JPhA...52W5003C}. These exact
results were extended to finite volume in \cite{2020PhRvX..10a1054B}
with a proof that applies to systems including Heisenberg spin chains,
the Lieb-Liniger interacting Bose gas, and integrable quantum field
theories with diagonal scattering. In \cite{sajat-algebraic-currents}
an algebraic construction was given for the current operators of the
integrable spin chains, which led to an alternative rigorous proof
of their mean values.

Recently \cite{2020ScPP....8...16P} it was discovered that these
exact results are connected to the so-called long range deformations
of integrable spin chains which emerged in the context of the AdS/CFT
correspondence \cite{2004PhR...405....1B,2008JSMTE..11L.001B,2009JPhA...42B5205B}.
These long range spin chains are obtained as a one-parameter deformation
of ordinary short-range spin chains (such as the XXZ model) and preserve
integrability to each finite order in the deformation parameter $g$,
with the interaction range growing order-by-order. Strictly speaking,
these deformations are only defined in infinite volume. In the work
\cite{2020JHEP...03..092P} it was demonstrated that these long range
deformations have a deep connection to $T\bar{T}$-deformations of
integrable quantum field theory \cite{2017NuPhB.915..363S,2019JHEP...09..052C}
(see also \cite{1991NuPhB.358..524Z,2000NuPhB.578..527M}), by sharing
the same algebraic origin which allowed the proof of factorisation
for the expectation values of the operators which trigger the deformation
of the spin chain.

It turns out that the existence of the exact formulae for the current
expectation values is implied by the following observation made in
\cite{2020ScPP....8...16P}: for each generalised current there is
a long range deformation such that the given current operator itself
is the leading perturbing operator. This implies that the perturbation
of the spin chain by the generalised current operator is integrable
to the leading order in the deformation parameter $g$, but integrability
is generally expected to be broken at higher orders since to maintain
it necessitates the inclusion of progressively longer and longer range
interaction terms at higher orders of $g$. We call this breaking
of integrability 'weak' in distinction to 'strong' breaking of integrability
which already happens at the first order of perturbation, and where integrability cannot be maintained by improving the perturbing operator order-by-order in $g$. {At present it is believed that the only two possibilities for weak integrability breaking are given by the current- and $T\bar{T}$-like deformations, see \cite{2020ScPP....8...16P,2020JHEP...03..092P,2004PhR...405....1B,2008JSMTE..11L.001B,2009JPhA...42B5205B}.}

{The concept of weak integrability breaking is novel and it opens up new questions about the physical behaviour of integrable and nearly integrable models. Earlier works did not make a distinction between different forms of integrability breaking (for a recent review see \cite{2021arXiv210311997B})\footnote{{We note that while Ref. \cite{2021arXiv210311997B} also uses the expression ``weak integrability breaking'', it is used in a very different way from the present work, since there it simply means perturbation with small coupling constant, and does not refer to a special class of perturbing operators.}}, and the physical consequences of the different types of perturbations started to emerge only recently. For example the work
\cite{doyon-weak-breaking} studied integrability breaking within the framework of Generalised Hydrodynamics, where it was also
found that the perturbations by the current operators do not break
integrability at the leading order: while a generic perturbation with
coupling $\lambda$ is expected to thermalise the system on a time
scale of $T\sim\lambda^{-2}$, this effect is missing for the current
operators on the same time scale. We also note the recent work \cite{2021arXiv210503326D} which argues that the perturbations with weak integrability breaking actually span the tangent of space of integrable models within the full space of local many-body models. This is an alternative explanation for the existence of weak integrability breaking: it arises as an effect of following the ``tangent line'' instead of remaining within the submanifold of integrable models, thus leading to much weaker effects than a perturbation which is ``orthogonal'' to the tangent space.}

The onset of non-integrable a.k.a. chaotic behaviour in many-body
systems can be investigated via the statistics of energy spectra \cite{2004PhRvB..69e4403R,2010PhRvA..82a1604R,2010PhRvE..81c6206S,2010PhRvE..82c1130S,2010JSMTE..07..013B,2016AdPhy..65..239D,2015PhRvE..91a2144B,2018PhRvE..98b2204B,2020arXiv200812782H};
in some cases even analytical results can be obtained \cite{2018PhRvL.121z4101B}.
For integrable systems the level spacing distribution is known to follow Poissonian statistics, while for non-integrable case random matrix theory predicts the so-called Wigner-Dyson distribution whose precise form depends
on the relevant random matrix ensemble. {In a finite volume, the transition between the two distributions is a smooth crossover, with the crossover coupling going to zero in the thermodynamic limit. For systems without a spectral gap, the crossover coupling scales as a (negative) power of the volume, with the exponent depending whether the local degrees of freedom are interacting or not \cite{2014NJPh...16i3016M}; for the case with interactions, the behaviour was determined to be $L^{-3}$, irrespective of the spatial dimensionality of the system.}

In this short paper we set out to investigate the weak breaking of
integrability by exploring the level spacing statistics for the case
of the spin-1/2 XXZ spin chain. We compare a known strong integrability
breaking term which introduces {next-to-nearest-neighbour} 
interactions to the weak integrability breaking perturbation provided by the lowest non-trivial generalised current. We start by presenting the Hamiltonian and the perturbations in Section \ref{sec:The-Hamiltonian-and}, and then
turn to the results for the level spacing statistics in Section \ref{sec:Level-spacing-distribution},
where we examine the crossover between the Poissonian distribution
characteristic for integrable systems to the Wigner-Dyson distribution
signalling the breaking of integrability, as a function of the integrability breaking coupling and the volume, {and investigate the dependence of the crossover coupling on the volume}. We present our conclusions in Section
\ref{sec:Conclusions}.

\section{The Hamiltonian and its perturbations \label{sec:The-Hamiltonian-and}}

\subsection{The XXZ spin chain and the current operator}

The spin-1/2 XXZ spin chain of length $L$ is defined by the Hamiltonian
\begin{equation}
H_{XXZ}=\sum_{i=1}^{L}\left[s_{i}^{x}s_{i+1}^{x}+s_{i}^{y}s_{i+1}^{y}+\Delta s_{i}^{z}s_{i+1}^{z}\right]\label{eq:XXZ_Hamiltonian}
\end{equation}
where $s^{j}=\frac{1}{2}\sigma^{j}$ are the spin operators with $\sigma^{j}$
denoting the Pauli matrices
\begin{equation}
\sigma^{x}=\begin{bmatrix}0 & 1\\
1 & 0
\end{bmatrix}\quad\sigma^{y}=\begin{bmatrix}0 & -i\\
i & 0
\end{bmatrix}\quad\sigma^{z}=\begin{bmatrix}1 & 0\\
0 & -1
\end{bmatrix}
\end{equation}
and periodic boundary conditions $s_{L+1}^{a}\equiv s_{1}^{a}$. The
model has three phases controlled by the anisotropy parameter $\Delta$:
for $-1<\Delta<1$ the spectrum is gapless, while for $\Delta>1$
and $\Delta<-1$ there is a non-zero gap. {The two massive phases are physically different: for $\Delta<-1$ the system is in a ferromagnetic Ising phase, while $\Delta>1$ corresponds to an antiferromagnetic Ising phase.} The boundary points $\Delta=\pm1$ are special as the $U(1)$ symmetry of the theory generated by the conserved charge
\begin{equation}
S^{z}=\sum_{i=1}^{L}s_{i}^{z}
\end{equation}
is enhanced to $SU(2)$.

The XXZ chain is integrable: there exists a family of local conserved
quantities
\begin{equation}
Q_{n}=\sum_{l=1}^{L}q_{n,l}
\end{equation}
where $n=1,\dots,L$ and the charge density is supported on exactly
$n$ neighbouring sites. Their conservation can be expressed as a
continuity equation
\begin{equation}
j_{n,l+1}-j_{n,l}=i\left[q_{n,l},H_{XXZ}\right]
\end{equation}
where $j_{n,l}$ is the conserved current corresponding to the charge
$Q_{n}$.

The first two charges are $Q_{1}=S^{z}$, and $Q_{2}$ which is nothing
else but the Hamiltonian \eqref{eq:XXZ_Hamiltonian}. These charges
are related to the $U(1)$ symmetry and time translations, and are
generally present in all the models considered in this work, including
the non-integrable ones.

The first nontrivial charge corresponding to integrability of the
XXZ Hamiltonian is
\begin{align}
Q_{3} & =\sum_{l=1}^{L}q_{3,l}\\
 & q_{3,l}=s_{l-1}^{x}s_{l}^{z}s_{l+1}^{y}-s_{l-1}^{y}s_{l}^{z}s_{l+1}^{x}+\Delta\left(-s_{l-1}^{z}s_{l}^{x}s_{l+1}^{y}+s_{l-1}^{z}s_{l}^{y}s_{l+1}^{x}-s_{l-1}^{x}s_{l}^{y}s_{l+1}^{z}+s_{l-1}^{y}s_{l}^{x}s_{l+1}^{z}\right)\nonumber
\end{align}
with the corresponding current given by
\begin{align*}
j_{3,l}=-\frac{1}{2}\bigg[ & 2\Delta\big(s_{l-2}^{x}s_{l-1}^{y}s_{l}^{x}s_{l+1}^{y}+s_{l-2}^{x}s_{l-1}^{z}s_{l}^{x}s_{l+1}^{z}+s_{l-2}^{y}s_{l-1}^{x}s_{l}^{y}s_{l+1}^{x}+s_{l-2}^{y}s_{l-1}^{z}s_{l}^{y}s_{l+1}^{z}\\
 & +s_{l-2}^{z}s_{l-1}^{x}s_{l}^{z}s_{l+1}^{x}+s_{l-2}^{z}s_{l-1}^{y}s_{l}^{z}s_{l+1}^{y}-s_{l-2}^{x}s_{l-1}^{y}s_{l}^{y}s_{l+1}^{x}-s_{l-2}^{y}s_{l-1}^{x}s_{l}^{x}s_{l+1}^{y}\big)\\
 & -2\left(s_{l-2}^{x}s_{l-1}^{z}s_{l}^{z}s_{l+1}^{x}+s_{l-2}^{y}s_{l-1}^{z}s_{l}^{z}s_{l+1}^{y}\right)-2\Delta^{2}\left(s_{l-2}^{z}s_{l-1}^{x}s_{l}^{x}s_{l+1}^{z}+s_{l-2}^{z}s_{l-1}^{y}s_{l}^{y}s_{l+1}^{z}\right)\bigg]\\
 & -\frac{1+\Delta^{2}}{4}\left(s_{l-1}^{x}s_{l}^{x}+s_{l-1}^{y}s_{l}^{y}\right)-\frac{\Delta}{2}s_{l-1}^{z}s_{l}^{z}\,.
\end{align*}

\subsection{Perturbations, norms and effective coupling}

We are interested in perturbing the XXZ chain by the current
\begin{equation}
H_{J}=H_{XXZ}+g_{3}J\,,
\end{equation}
where
\begin{equation}
J=\sum_{l=1}^{L}j_{3,l}\label{eq:Jdef}
\end{equation}
which is supposed to break integrability only at higher order. As
a benchmark to the strength of integrability breaking, we also consider
a perturbation breaking integrability by the next-to-nearest-neighbor
interaction (NNNI) term
\begin{equation}
\op_{NNNI}=\sum_{i=1}^{L}s_{i}^{z}s_{i+2}^{z}\,,\label{eq:NNNIdef}
\end{equation}
which leads to the Hamiltonian
\begin{equation}
H_{NNNI}=H_{XXZ}+g_{N}\sum_{i=1}^{L}s_{i}^{z}s_{i+2}^{z}=\sum_{i=1}^{L}\left[s_{i}^{x}s_{i+1}^{x}+s_{i}^{y}s_{i+1}^{y}+\Delta s_{i}^{z}s_{i+1}^{z}+g_{N}s_{i}^{z}s_{i+2}^{z}\right]\,.
\end{equation}
{We remark that integrability breaking also affects transport properties, which was investigated in \cite{2006PhRvL..96f7202J}  exactly for the above NNNI perturbation of the spin-1/2 XXZ chain.}

To compare the strength of integrability breaking, we introduce the
effective coupling $g_{\text{eff}}$ of an operator $\op$ as
\begin{equation}
g_{\text{eff}}=g_{\mathcal{O}}n_{\mathcal{O}}
\end{equation}
where $g_{\mathcal{O}}$ is the coupling appearing in the Hamiltonian
$H=H_{0}+g_{\mathcal{O}}\op$, and $n$ is defined from the norm of
the operator 
\begin{equation}
\left\Vert \op\right\Vert _{2}=\sup_{\Vert x\Vert=1}\Vert \mathcal{O}x\Vert
\end{equation}
as the coefficient $n$ of the leading asymptotic term
\begin{equation}
||\op||_{2}=n_{\mathcal{O}}L+...\:,
\end{equation}
which is linear in the volume $L$ due to the expression of $\op$
as the translation invariant sum of localised terms (\ref{eq:Jdef},\ref{eq:NNNIdef}).
{The advantage in parameterising the strength of the perturbation with $g_{\text{eff}}$ is that its value is invariant under a rescaling of the perturbing operator, and it also facilitates the comparison of the strengths of different perturbing operators}.

\begin{figure}[H]
\centering{}\includegraphics[width=0.8\linewidth]{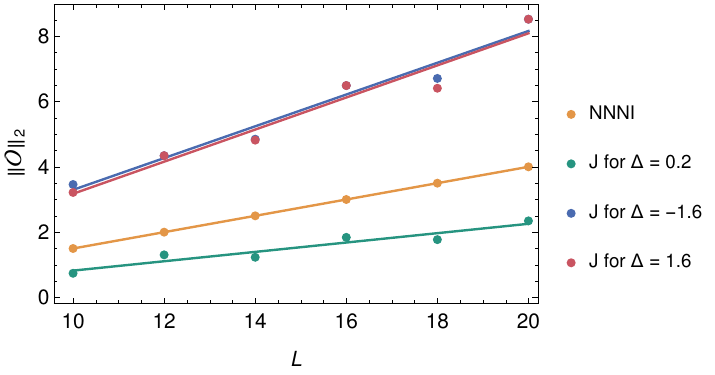} \caption{\label{fig:Norm-of-} Norm of $J$ and $\op_{NNNI}$ as a function
of the system size $L$ in the three phases.}
\end{figure}

The norm of $J$ depends on $\Delta$, while that of $\op_{NNNI}$
does not; some explicit values are shown in Table \ref{tab:The-norms-of}. {Since the perturbing operators are given as translation invariant sums over localised interaction terms, their norm is expected to be extensive in the volume. Indeed, }
as illustrated in Fig. \ref{fig:Norm-of-}, their norms change linearly
with the volume, apart from some fluctuations for $J$ which are due
to the fact that $L$ is changed in steps of $2$, while the one-site
term $j_{3,l}$ is localised on $4$ sites. The coefficient $n_{\mathcal{O}}$
can be computed by fitting a linear function to the norm values as
a function of $L$, and is shown in the last column of Table \ref{tab:The-norms-of}.

\begin{table}[H]
\begin{centering}
\begin{tabular}{@{}l@{}lllllll@{}l@{}}
\toprule
 & $L$  & $10$  & $12$  & $14$  & $16$  & $18$  & $20$  & ~ $n_{\mathcal{O}}$\tabularnewline
\midrule
\multirow{3}{*}{$\left\Vert J\right\Vert _{2}$} & $\Delta=0.2$  & $0.742$  & $1.306$  & $1.232$  & $1.841$  & $1.772$  & $2.346$  & ~ $0.143$\tabularnewline
\cmidrule{2-9} \cmidrule{3-9} \cmidrule{4-9} \cmidrule{5-9} \cmidrule{6-9} \cmidrule{7-9} \cmidrule{8-9} \cmidrule{9-9}
 & $\Delta=-1.2$  & $2.376$  & $3.115$  & $3.357$  & $4.614$  & $4.726$  & $6.023$  & ~ $0.348$\tabularnewline
\cmidrule{2-9} \cmidrule{3-9} \cmidrule{4-9} \cmidrule{5-9} \cmidrule{6-9} \cmidrule{7-9} \cmidrule{8-9} \cmidrule{9-9}
 & $\Delta=1.2$  & $2.274$  & $3.115$  & $3.455$  & $4.614$  & $4.412$  & $6.023$  & ~ $0.340$\tabularnewline
\midrule
$\left\Vert \op_{NNNI}\right\Vert _{2}$  &  & $1.500$  & $2.000$  & $2.500$  & $3.000$  & $3.500$  & $4.000$  & ~ $0.250$\tabularnewline
\bottomrule
\end{tabular}
\par\end{centering}
\caption{\label{tab:The-norms-of} The norms of the perturbing operators $J$
and $\op_{NNNI}$ }
\end{table}

\section{Level spacing distribution in the XXZ chain and its perturbations
\label{sec:Level-spacing-distribution}}

The level spacings of a given system described by a Hamiltonian $H$
with eigenvalues $\lambda_{i}$ are defined as the differences $S_{i}=\lambda_{i+1}-\lambda_{i}$ between eigenvalues ordered as a monotonically increasing sequence $\lambda_1\leq\lambda_2\leq\dots$. The distribution of normalised level spacings $s_{i}=S_{i}/\bar{S}$ (where $\bar{S}$ is the mean level spacing) is called the level
spacing distribution $P(s)$. For integrable systems, the
level spacing distribution is exponential:
\begin{equation}
P(s)_{I}=e^{-s}\,,
\end{equation}
while for non-integrable systems it is given by the Wigner-Dyson distribution,
which takes the following form for the orthogonal Gaussian ensemble\footnote{Due to the fact that the Hamiltonians considered here are real and
symmetric, it is the orthogonal Gaussian ensemble which is relevant
here.}
\begin{equation}
P(s)_{NI}=\frac{\pi}{2}s\e{-\frac{\pi}{4}s^{2}}.
\end{equation}
Therefore, the level spacing distribution is an explicit indicator
of integrability and its breaking. The above predictions follow from
random matrix theory. For an integrable Hamiltonian, the different
levels do not interact due to the presence of higher conserved charges,
and so the distribution of eigenvalues is a Poissonian one, leading
to exponential distribution of the normalised level spacings. Breaking
integrability results in level repulsion, and so small level spacings
are suppressed. When considering the spectrum as a function of a parameter
such as volume, integrability implies that levels approaching each other as a function of volume simply cross, while in the non-integrable
case they avoid each other due to level repulsion. In the limit of infinite
matrix size, the level spacing distribution changes suddenly from
exponential to Wigner-Dyson for any non-zero value of the integrability
breaking coupling $g$.

For a spin chain of finite length, however, the Hilbert space is finite
dimensional, and so the level spacing distribution is a continuous
function of the coupling, with the transition becoming sharper for
larger volumes \cite{2004PhRvB..69e4403R,2010JSMTE..07..013B}. In
addition, when constructing the level spacing from the full spectrum
it is found to deviate from the random matrix prediction due to the
structure dictated by quasi-particle excitations, which is in turn due to
the locality of the Hamiltonian. This problem can be solved by constructing
the level spacing distribution from the middle part of the spectrum,
for which we take the middle two-thirds of the computed levels. Furthermore,
to get rid of degeneracies corresponding to trivial symmetries \cite{2003PhRvB..68e2510K,2005JPSJ...74.1992K}
the level spacing distribution is extracted from a sector with total
momentum zero, even spatial parity and a fixed {(non-zero)} $S_{z}$ value\footnote{{Note that the $S_{z}=0$ sector has an additional spin-flip symmetry.}}. In the examples below we present the results obtained for $S_{z}=2$; similar results were obtained for $S_{z}=1$ and $3$.

\subsection{The integrable XXZ chain}

\begin{figure}[H]
\centering \includegraphics[width=0.34\linewidth]{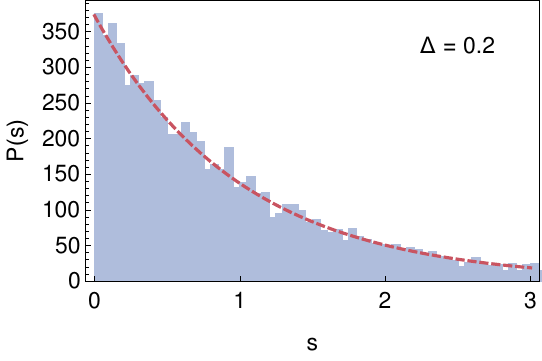}\includegraphics[width=0.34\linewidth]{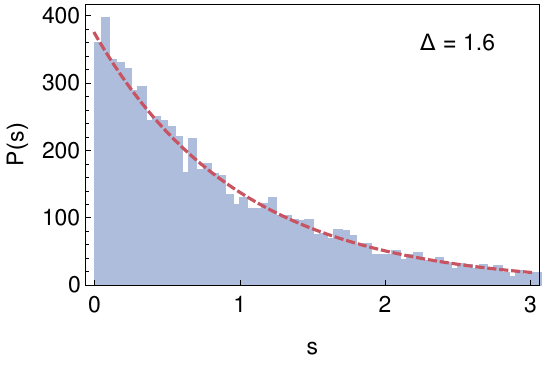}\includegraphics[width=0.34\linewidth]{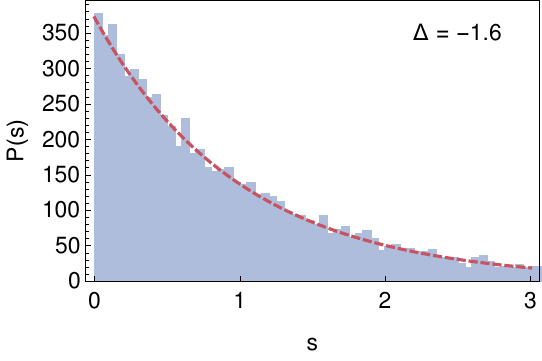}
\caption{Level spacing statistics of the XXZ spin chain of length $L=22$ for
(left to right) $\Delta=0.2,1.6,-1.6$ with the exponential distribution
fitted (dashed red line).}
\label{xxzfit}
\end{figure}

The level spacing statistics of the XXZ chain of length $L=22$ in
the different phases is shown in figure \ref{xxzfit}. {It can be described very well by an exponential curve, with the overall normalisation of the distribution as the only fitting parameter.}

\subsection{Perturbed system and crossover from Poisson to Wigner-Dyson statistics}

\begin{figure}[H]
\centering \includegraphics[width=0.35\linewidth]{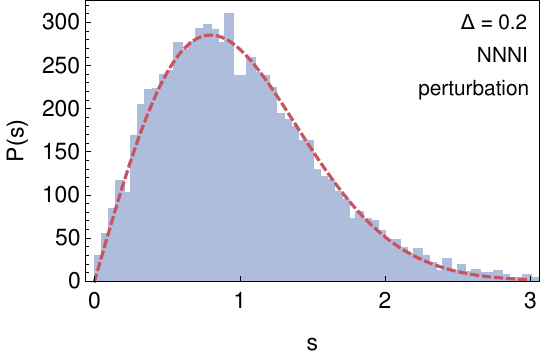}\hspace{1cm}\includegraphics[width=0.35\linewidth]{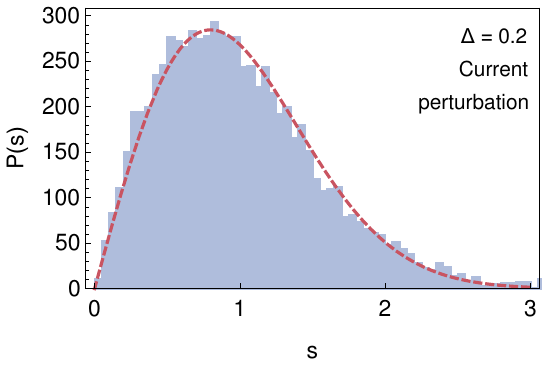}
\caption{{A typical level spacing distribution of a non-integrable system with
the Wigner-Dyson distribution fitted (red dashed line). The left panel shows
the case of $H_{NNNI}$ with $L=22$ $\Delta=0.2$ and $g_{\text{eff}}=0.1$, while the right panel belongs to $H_{J}$ with $L=22$, $\Delta=0.2$ and $g_{\text{eff}}=0.42$.}}
\label{wdfit}
\end{figure}

{Switching on a suitably large value for the integrability breaking coupling the level spacing distribution is changed into the Wigner-Dyson statistics characteristic of quantum chaos, as illustrated in Fig. \ref{wdfit}. Again, the only fitting parameter is the overall normalisation of the distribution.}

{In a finite volume, varying the strength of the integrability breaking coupling leads to a crossover between the exponential and Wigner-Dyson distributions. The crossover can be quantified by determining the position of the maximum of the level-spacing distribution, which moves from the origin to the position $\sqrt{2/\pi}$ characteristic for the Wigner-Dyson distribution.}

\begin{figure}[H]
\centering 
\begin{subfigure}[b]{0.95\textwidth}
\centering 
\includegraphics[width=.95\textwidth]{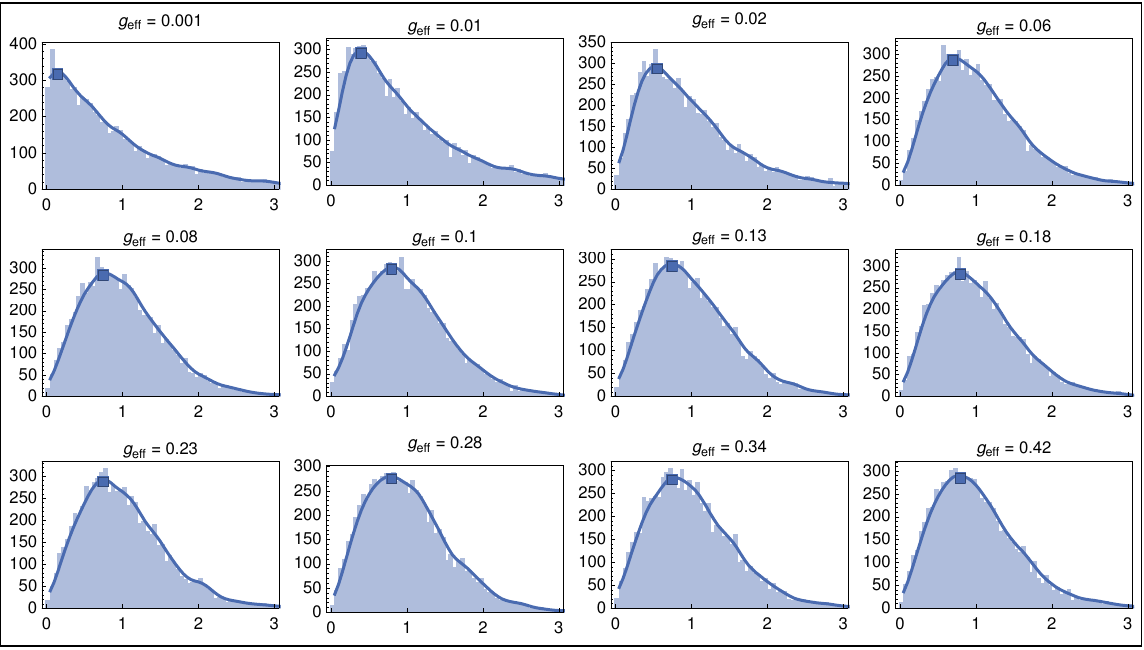}
\caption{NNNI perturbation}
\end{subfigure}
\begin{subfigure}[b]{0.95\textwidth}
\centering 
\includegraphics[width=.95\textwidth]{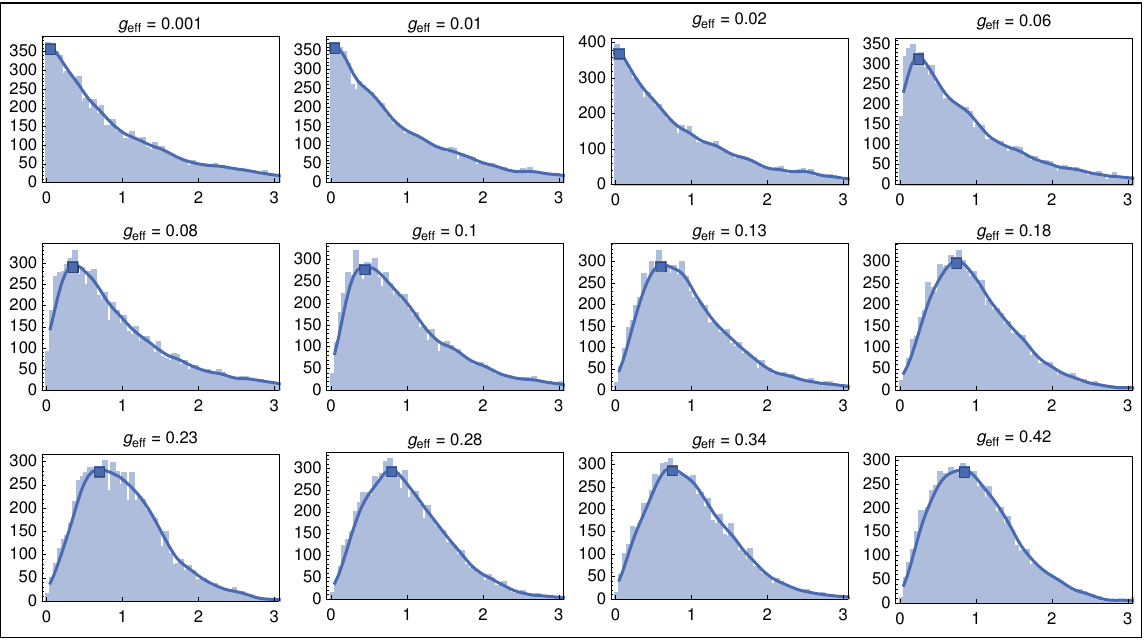}
\caption{Current perturbation}
\end{subfigure}
\caption{\label{peaks-nnni-and-j} Determining the dependence of the peak position on $g_\mathrm{eff}$ for the NNNI (a) and current (b) perturbations for $L=22$, $\Delta=0.2$. {The blue solid lines mark the result of the Gaussian filtering and} the blue markers denote the extracted peak positions.}
\end{figure}

{The normalised level spacings were sorted into bins of width $0.15$ for $L=16,18,20$, and of width $0.1$ for $L=22,24$. The resulting distribution was then smoothed by applying a Gaussian filter of kernel radius $r$ to the raw histogram to suppress fluctuations due to finite bin size, with the choice $r = 6$ for lengths $L = 16$ and $18$, and $r = 4$ for longer chains\footnote{{For chains with length smaller than $16$ there are simply not enough level spacings to yield useful statistics.}}. The determination of the position of the maximum of the level spacing distribution is illustrated in Fig. \ref{peaks-nnni-and-j} for a chain of length $L=22$.}

{Note that the crossover happens at smaller couplings for the NNNI than for the current perturbation, indicating the difference in the `strength' of integrability breaking; however, the decisive evidence eventually comes from the volume dependence considered in Subsection \ref{volume-dependence}.}

\begin{figure}[H]
\centering \includegraphics[width=0.8\linewidth]{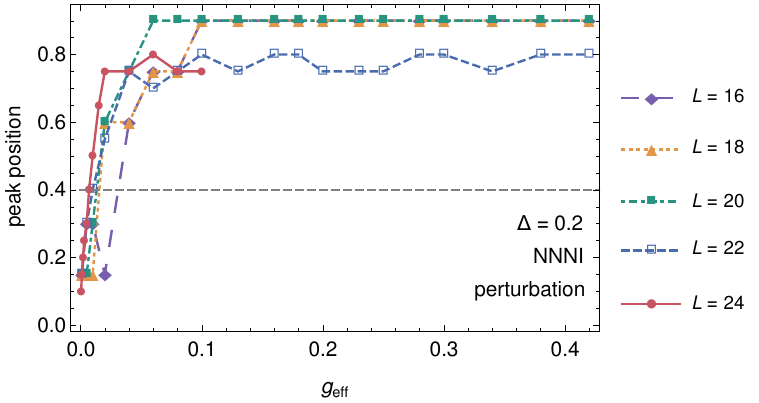} \caption{{The position of the peak of the level spacing distribution corresponding to $H_{NNNI}$ for $\Delta=0.2$ as a function of the effective coupling $g_\text{eff}$ for different chain lengths $L$. The dashed grey line marks $x_0$, where $2x_0 = \sqrt{2/\pi}$ is the position of the maximum of the exact Wigner-Dyson distribution. The transition from Poissonian to Wigner-Dyson statistics is faster for longer chains, as expected.}}
\label{nnni-d02}
\end{figure}

\begin{figure}[H]
\centering \includegraphics[width=0.8\linewidth]{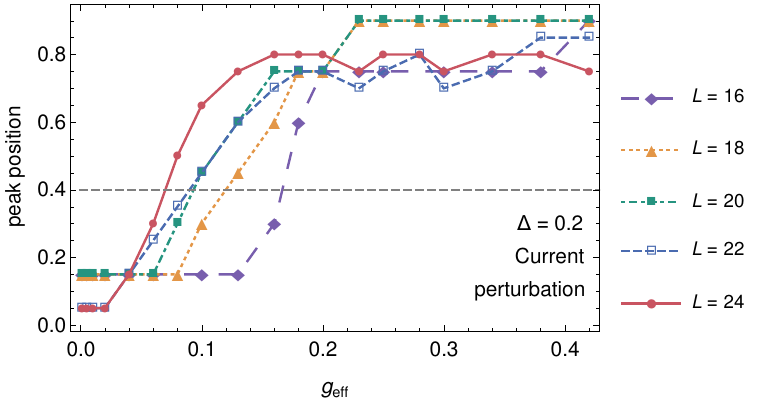} \caption{{The position of the peak of the level spacing distribution corresponding to $H_{J}$ for $\Delta=0.2$ as a function of the effective coupling $g_\text{eff}$ for different chain lengths $L$. The dashed grey line marks $x_0$, where $2x_0 = \sqrt{2/\pi}$ is the position of the maximum of the exact Wigner-Dyson distribution. The transition from Poissonian to Wigner-Dyson statistics is faster for longer chains, as expected.}}
\label{curr-d02}
\end{figure}

{Results for the NNNI perturbation in the gapless phase can be seen in Fig. \ref{nnni-d02}, while for the current perturbation they are shown in Fig. \ref{curr-d02}. We remark that while the filtering facilitates the finding of the peak, the precision of its determined location is still limited by the bin size, leading to fluctuations in the determined peak positions which can be seen in the figures. As expected, the crossover occurs faster for
longer chains, and the data also show that it is markedly slower for the
current perturbation than for the NNNI case, supporting the idea
that the current perturbation only breaks integrability at higher orders.}

\subsection{Integrability breaking and perturbation theory \label{subsec:Integrability-breaking-and}}

\begin{figure}[H]
\centering \includegraphics[width=1\linewidth]{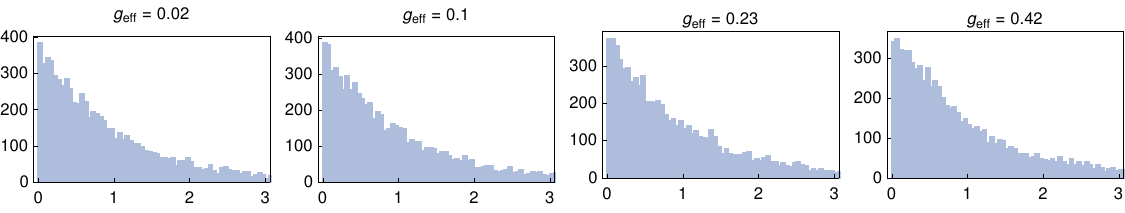}
\caption{\label{peaks-nnni-pert}Level spacing statistics of $H_{NNNI}$ as
computed from the spectrum obtained by first order perturbation theory
for the case $L=22$, $\Delta=0.2${, for different values of the coupling.}}
\end{figure}

To investigate the order of integrability breaking it is tempting
to try and construct the level spacing distribution for the spectrum
constructed from perturbation theory in the coupling $g$. However
it turns out that this is not possible. {Using simple first-order matrix perturbation theory to compute the spectrum of the perturbed Hamiltonian, and evaluating the resulting level spacing distribution demonstrates that it remains exponential even for the NNNI perturbation \eqref{eq:NNNIdef}, as shown in Fig. \ref{peaks-nnni-pert}}.

To understand the reason, consider how a level crossing is lifted
by a perturbation. In our model, the generic level crossing obtained
as a function of the volume $L$ (continued to real values) happens
between two levels. Let us consider {an effective Hamiltonian description of} a generic perturbation of the corresponding two level subsystem:
\begin{equation}
\begin{bmatrix}a_{1}(L)+g\Delta a_{1}(L) & g\epsilon(L)\\
g\epsilon(L) & a_{2}(L)-g\Delta a_{2}(L)
\end{bmatrix}\,,
\end{equation}
{where the unperturbed energy levels are given by $a_{1,2}(L)$, and the functions $\Delta a_{1,2}(L)$ and $\epsilon(L)$ parameterise a general integrability breaking perturbation with strength $g$, in the two-level subspace.}
For the unperturbed Hamiltonian $g=0$, the level crossing is located
at the volume $L_{0}$ given by $a_{1}(L_{0})=a_{2}(L_{0})$; all
that happens at first order is that the level crossing is shifted
to a location given to the location $L_{*}$ which satisfies $a_{1}(L_{*})+g\Delta a_{1}(L_{*})=a_{1}(L_{*})-g\Delta a_{2}(L_{*})$,
where $L_{*}$ can be computed as a series in $g$:
\begin{equation}
L_{*}(g)=L_{0}-g\frac{\Delta a_{1}(L_{0})-\Delta a_{1}(L_{0})}{a_{1}'(L_{0})-a_{2}'(L_{0})}+O\left(g^{2}\right)\,.
\end{equation}
Therefore first order perturbation theory does not introduce a repulsion
between levels, so the exponential distribution is unchanged.

{When extending the perturbative calculation of the level spacing statistics to second order, the resulting energy levels turn out to be numerically unstable and no meaningful statistics can be constructed.} For the two-level system above, an exact calculation of the energy levels shows that the off-diagonal term leads to level repulsion, which is responsible for changing the level spacing distribution. However, a perturbative evaluation of the two nearby energy levels results in
\begin{equation}
E_{1,2}=a_{1,2}(L)+g\Delta a_{1,2}(L)\pm g^{2}\frac{2\epsilon(L)^{2}}{a_{1}(L)-a_{2}(L)}+O\left(g^{3}\right)
\end{equation}
which is unstable in the vicinity of a level crossing due to the presence of the energy difference denominator. As a result, the order of integrability
breaking cannot be deduced by considering level spacing distribution
of the spectrum obtained in perturbation theory in the coupling $g$.

\subsection{Volume dependence of the crossover coupling} \label{volume-dependence}

For the system with the integrability breaking turned on, let's define the crossover coupling $g_{\text{cr}}$ as the value of the effective coupling $g_{\text{eff}}$ for which the maximum of the intermediate distribution is at $x_{0}$, where $2 x_{0}=\sqrt{2/\pi}$ is the position of the maximum of the exact Wigner-Dyson distribution. The dependence of the crossover coupling on the volume $L$, $g_{\text{cr}}(L)$ is expected to be a monotonically decreasing function. More precisely, in the gapless phase it is expected to have a power-like behaviour $g_{\text{cr}}\propto L^{-\alpha}$ \cite{2014NJPh...16i3016M,2014PhRvB..90g5152M}, while in the gapped phase finite volume corrections are expected to show exponential decay in the volume (cf. Subsection \ref{gapped-phase}).

{Note that since the transition is a smooth crossover, other definitions of $g_{\text{cr}}$ are also possible \cite{2014NJPh...16i3016M,2014PhRvB..90g5152M}; however, the finite size scaling is expected to be universal and therefore independent of these details. Indeed, in the following we recover the exponent $\alpha=3$ obtained previously for the NNNI perturbation in the gapless phase \cite{2014PhRvB..90g5152M}.}

\subsubsection{Gapless phase}

In the following we give results obtained by carrying out the above
described method for the NNNI and current perturbations in the gapless
phase, namely for $\Delta=0.2$.

\begin{figure}[H]
\centering \includegraphics[width=0.5\linewidth]{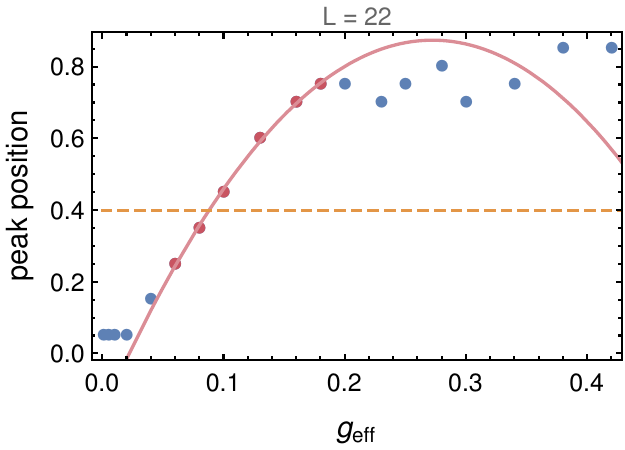}\includegraphics[width=0.5\linewidth]{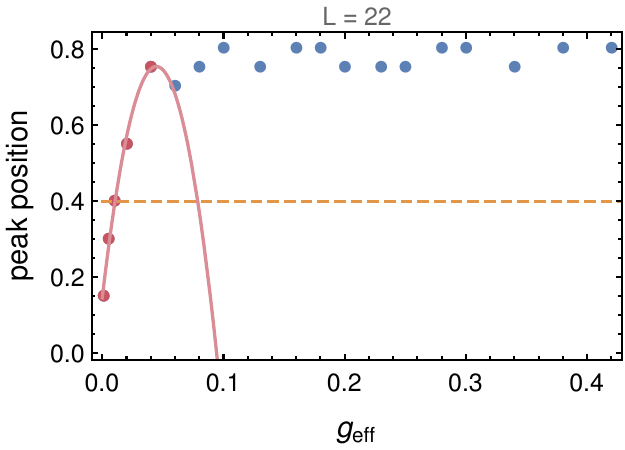}
\caption{\label{crossover-fit} Obtaining $g_{cr}$ from the maximum positions
for different effective couplings for $J$ (left) and NNNI perturbation
(right) in the gapless phase ($\Delta = 0.2$). The red solid line is the parabola fitted to the red markers
in the vicinity of $x_{0}$ (indicated by orange dashed line).}
\end{figure}

The obtained maximum positions can be seen in figure \ref{crossover-fit}
along with the parabola $f(g)$ fitted to the resulting points around
$x_{0}$. The crossover coupling is then obtained by solving $f(g_{cr})=x_{0}$.

\begin{figure}[H]
\centering \includegraphics[width=0.5\linewidth]{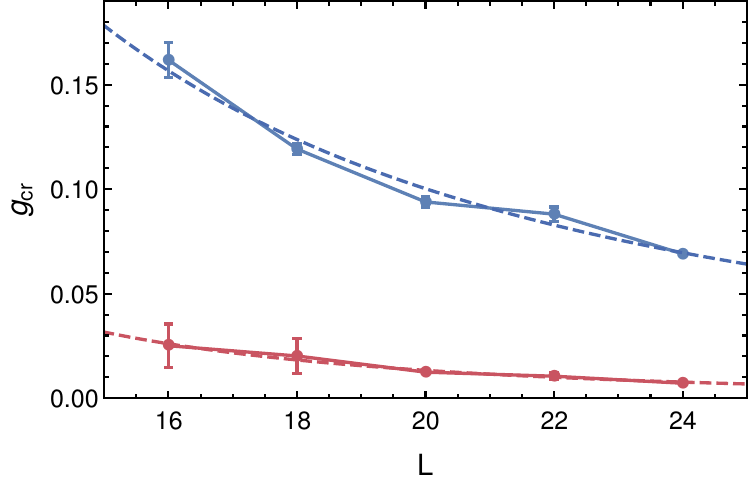}\includegraphics[width=0.5\linewidth]{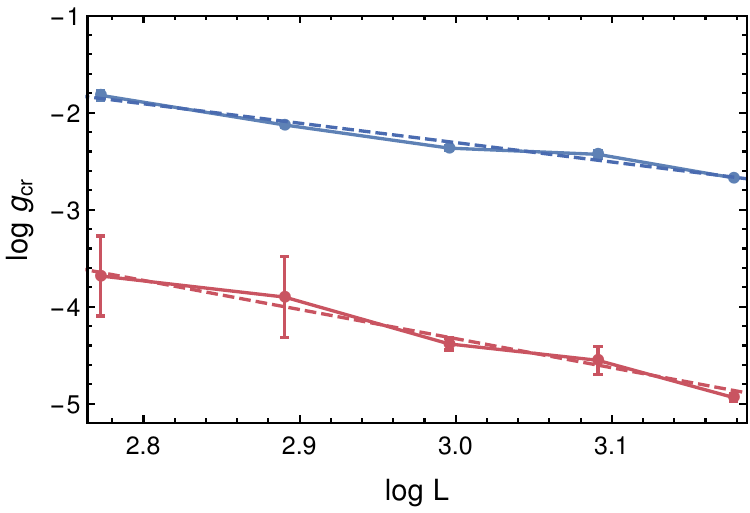}
\caption{\label{gcr} The crossover scale as a function of the volume for the
current (blue markers) and NNNI perturbations (red markers) on normal
(left) and log-log scale (right) as obtained in the gapless phase ($\Delta = 0.2$). Dashed lines correspond to the fitted $AL^{-3}$ (red) and $BL^{-2}$ (blue) respectively.}
\end{figure}

The $g_{cr}(L)$ functions obtained this way for both the NNNI and
current perturbations can be seen in figure \ref{gcr}, and decay
with a power of the volume $L$. Fitting a linear function $\log g_{cr}(L)=a+b\log L$
results in the exponents
\begin{align}
 & b_{J}=-1.99\pm0.18\nonumber\\
 & b_{NNNI}=-3.11\pm0.27
\end{align}
The value obtained for the NNNI case is consistent with the universal
exponent $-3$ claimed for integrability breaking in a gapless chain 
\cite{2014NJPh...16i3016M,2014PhRvB..90g5152M}, while the one obtained for the current perturbation is in agreement with the conjecture that the current perturbation breaks integrability at higher orders in perturbation
theory.

\subsubsection{Gapped phase}\label{gapped-phase}

In the gapped phase, finite size scaling is expected to be exponential
in volume, {since the presence of finite correlation length implies exponential decay of correlations. As a result, finite size effects on the spectrum (beyond its inevitable discretisation in finite volume) generally decay exponentially with the volume \cite{1986CMaPh.104..177L,1986CMaPh.105..153L}}. 

\begin{figure}[H]
\centering{}\includegraphics[width=0.5\linewidth]{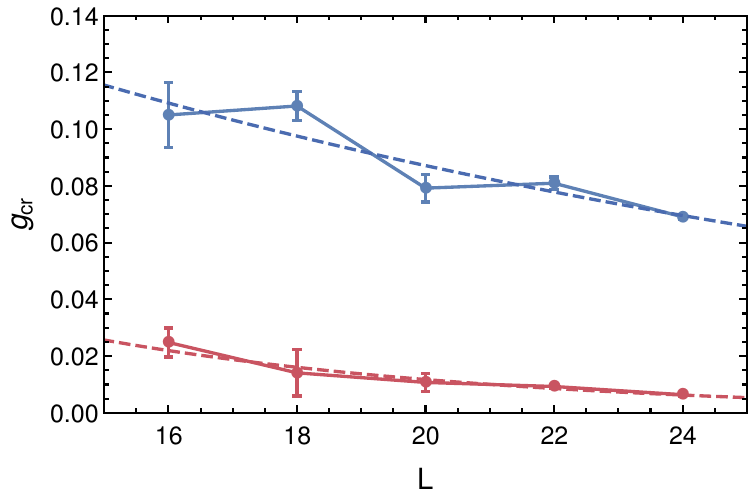}\includegraphics[width=0.5\linewidth]{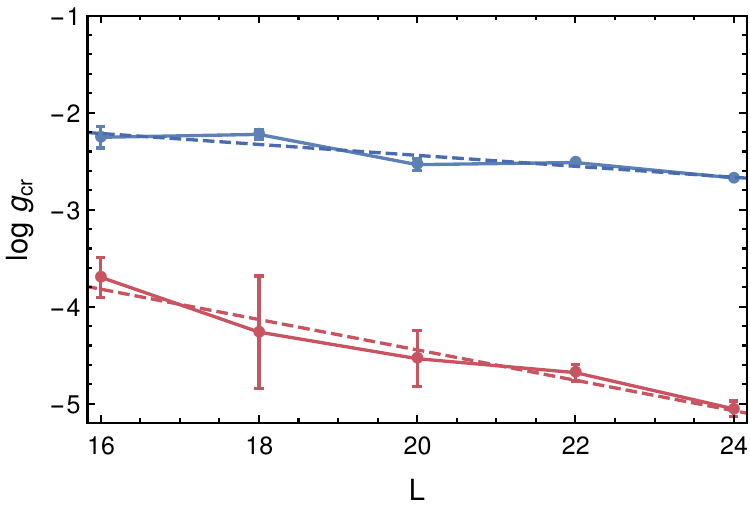}
\caption{\label{fig:The-crossover-scale-massive} The crossover scale as a
function of the volume for the current (blue markers) and NNNI perturbations
(red markers) for $\Delta=1.6$ (gapped phase, antiferromagnetic) on normal (left) and
log scale (right). Dashed lines correspond to the fitted exponentials.}
\end{figure}

\begin{figure}[H]
\centering{}\includegraphics[width=0.5\linewidth]{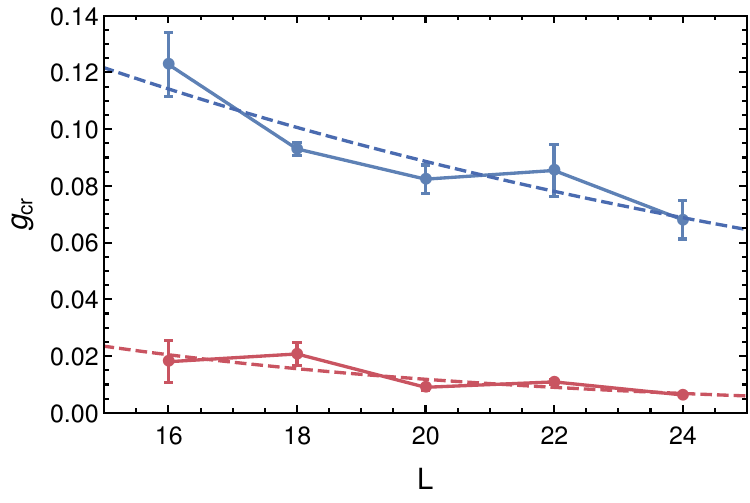}\includegraphics[width=0.5\linewidth]{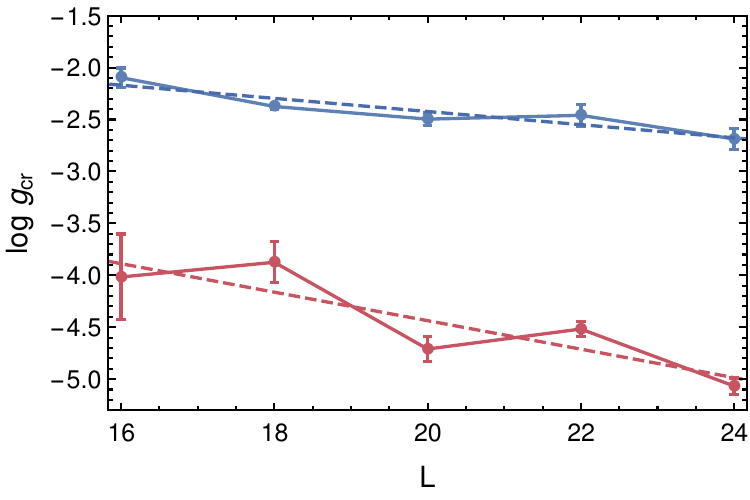}
\caption{\label{fig:The-crossover-scale-minus16}The crossover scale as a function of the volume for the current (blue markers) and NNNI perturbations
(red markers) for $\Delta=-1.6$ (gapped phase, ferromagnetic) on normal (left) and log scale (right). Dashed lines correspond to the fitted exponentials.}
\end{figure}

The crossover coupling as a function of the volume is shown
for $\Delta=1.6$ in Fig. \ref{fig:The-crossover-scale-massive},
and for $\Delta=-1.6$ in Fig. \ref{fig:The-crossover-scale-minus16}.
For the massive case, $\log g_{cr}(L)$ can be fitted with a linear
function $c+dL$, with the following values for the coefficients $d$
for $\Delta=1.6$:
\begin{align}
 & d_{J}=-0.056\pm0.014\nonumber\\
 & d_{NNNI}=-0.157\pm0.020
\end{align}
and for $\Delta=-1.6$:
\begin{align}
 & d_{J}=-0.063\pm0.014\nonumber\\
 & d_{NNNI}=-0.137\pm0.043
\end{align}
Again we see a marked difference between the strength of integrability
breaking for the two perturbations. The crossover coupling for the
current perturbation in any given volume is an order of magnitude
larger than for the NNNI perturbation, and the coefficient $d$ describing
its decay with the volume is also significantly smaller, again in
agreement with the conjecture that the current perturbation breaks
integrability at higher orders in perturbation theory.

\section{Conclusions \label{sec:Conclusions}}

In this work we examined the crossover from integrable to chaotic
behaviour for weak breaking of integrability, defined as a perturbation
of an integrable system which preserves integrability to first order
in perturbation theory. As an example we took the XXZ spin chain perturbed
by one of the currents that appear in the continuity equation for
the higher conserved charges which imply integrability, and compared it to a usual (``strong'') integrability breaking perturbation which was chosen to be a  next-to-nearest-neighbour interaction term (NNNI).

The tool we used was the evaluation of level spacing distribution
for a finite chain using exact diagonalisation. Since the extremal
parts of the spectrum have special properties due to locality of the
Hamiltonian, they were discarded with the middle two-thirds of levels
kept. We then quantified the crossover between the integrable Poissonian
and chaotic Wigner-Dyson statistics in the form of a crossover coupling. 
To facilitate the comparison between different operators, we rescaled their coupling constants by the operator norms (per unit volume), which also makes the values of the couplings independent of the choice of operator normalisation. 

The behaviour of the crossover coupling as a function of the volume
was found to be fully consistent with the suggestion in \cite{2020ScPP....8...16P}
that the current perturbation only breaks integrability in higher
order. For any fixed volume $L$  the crossover values of the rescaled couplings were significantly higher for the current perturbation compared to the NNNI case, and
their decay with the volume was also significantly slower. In particular,
for the gapless case while the crossover value of the NNNI perturbation
follows the $1/L^{3}$ law found previously in \cite{2014PhRvB..90g5152M},
for the current perturbation we found a $1/L^{2}$ decay. In the massive
regime the crossover coupling decreases exponentially with the volume,
but again the decay for the current perturbation was significantly
slower than for the NNNI. {We note that this difference in the volume dependence is unaffected by the rescaling with the operator norms.}

{The slower finite-volume crossover between exponential and Dyson-Wigner distribution can be interpreted as a delayed onset of quantum chaos. The notion that there can be weaker and stronger versions of quantum chaotic behaviour has recently appeared in a different context related to out-of-time-ordered correlators \cite{weak-quantum-chaos}. However, the class of weak integrability breaking perturbations considered here define a concept of ``weak quantum chaos'' different from the one in \cite{weak-quantum-chaos}, where instead it was related to the finite dimensionality of the local Hilbert spaces.}

Since the $1/L^{3}$ law was claimed universal
for strong integrability breaking in gapless models \cite{2014NJPh...16i3016M,2014PhRvB..90g5152M} (albeit
without analytic support), it is tempting to speculate that the exponent eventually depends on the order at which integrability is broken.
However, as already known and also argued for in Subsection \ref{subsec:Integrability-breaking-and},
perturbation theory cannot be applied to examine the crossover in
the level spacing distribution if the level crossings resolved by
the integrability breaking perturbation are generic i.e. involve only
two levels. It is interesting to note that in quantum field theories
obtained as perturbation of conformal field theories the crossover
to the Wigner-Dyson behaviour takes place already in the perturbative
regime \cite{2020arXiv201208505S}. The essential difference with
the case considered here is that for the conformal spectrum the levels
are generically multiply degenerate. 

To sum up, our results strongly support the observations in \cite{2020ScPP....8...16P,doyon-weak-breaking} that perturbations of spin chains by generalised currents correspond to a weak integrability breaking which only happens at higher order in the perturbing coupling. {An interesting question left open is to clarify the dependence of the finite volume crossover behaviour, especially the exponent appearing in the gapless case, on the order of integrability breaking.}

\subsection*{Acknowledgements}

This work was partially supported by the National Research, Development
and Innovation Office (NKFIH) under the research grant K-16 No. 119204,
and also by the Fund TKP2020 IES (Grant No. BME-IE-NAT), under the
auspices of the Ministry for Innovation and Technology. G. T. was
also supported by the the National Research, Development and Innovation
Office (NKFIH) via the Hungarian Quantum Technology National Excellence
Program, project no. 2017-1.2.1-NKP- 2017-00001.

\providecommand{\href}[2]{#2}\begingroup\raggedright\endgroup

\end{document}